\documentclass[letterpaper, 10 pt, conference]{ieeeconf}  
\IEEEoverridecommandlockouts                          
\usepackage{graphics}
\usepackage{multirow}
\usepackage{amsmath} 
\usepackage{amssymb}  
\usepackage{graphicx}
\usepackage[normalem]{ulem}
\usepackage{algorithm}
\usepackage{algpseudocode}
\usepackage{cite}
\usepackage{soul}
\usepackage{comment}
\usepackage{tikz}
\usepackage{pgfplots}
\usepackage{pgfplotstable}
\pgfplotsset{compat=newest,compat/show suggested version=false}
\usepackage{graphviz}
\usetikzlibrary{math}
\usetikzlibrary{shapes,arrows}
\usetikzlibrary{arrows.meta}
\tikzset{>={Latex[width=2.5mm,length=2.5mm]}}
\usetikzlibrary{positioning}
\tikzstyle{block}=[draw opacity=0.7,line width=1.4cm]
\tikzset{arrow_e/.style = {->,> = latex'}}

\newcommand{\vect}[1]{\mathbf{#1}}
\newcommand{\real}{{\mathbb{R}}}

\newcommand{\boxend}{\hfill \ensuremath{\Box}}

\newtheorem{rem}{Remark}[section]

\newtheorem{lem}{Lemma}[section]
\newtheorem{defn}{Definition}

\allowdisplaybreaks

\title{\LARGE \bf
FORWARD: Feasibility Oriented Random-Walk Inspired Algorithm \\ for Radial Reconfiguration in Distribution Networks
}
\author{
\\Joan Vendrell, Russell Bent and Solmaz Kia, \emph{Senior Member, IEEE}
\thanks{The first and third authors are with the Mechanical and Aerospace Engineering department,
        University of California Irvine, 
        {\tt\small jvendrel}, {\tt\small solmaz@uci.edu}. The second author is with the Theoretical Division,
        Los Alamos National Lab
        {\tt\small rbent@lanl.edu}. This work was supported by UCI-LANL Fellowship.}%
}
\allowdisplaybreaks

\usepackage{placeins}
\begin{document}
\raggedbottom
\maketitle
\begin{abstract}
We consider an optimal flow distribution problem in which the goal is to find a radial configuration that minimizes resistance-induced quadratic distribution costs while ensuring delivery of inputs from multiple sources to all sinks to meet their demands. This problem has critical applications in various distribution systems, such as electricity, where efficient energy flow is crucial for both economic and environmental reasons. Due to its complexity, finding an optimal solution is computationally challenging and NP-hard. In this paper, we propose a novel algorithm called \textsf{FORWARD}, which leverages graph theory to efficiently identify feasible configurations in polynomial time. By drawing parallels with random walk processes on electricity networks, our method simplifies the search space, significantly reducing computational effort while maintaining performance. The \textsf{FORWARD} algorithm employs a combination of network preprocessing, intelligent partitioning, and strategic sampling to construct radial configurations that meet flow requirements, finding a feasible solution in polynomial time. Numerical experiments demonstrate the effectiveness of our approach, highlighting its potential for real-world applications in optimizing distribution networks.
\end{abstract}
\section{Introduction}
\label{sec::intro}
\begin{figure}[t]
    \centering
    \begin{tabular}{c}
       \includegraphics[width=0.9\linewidth,height=0.45\linewidth]{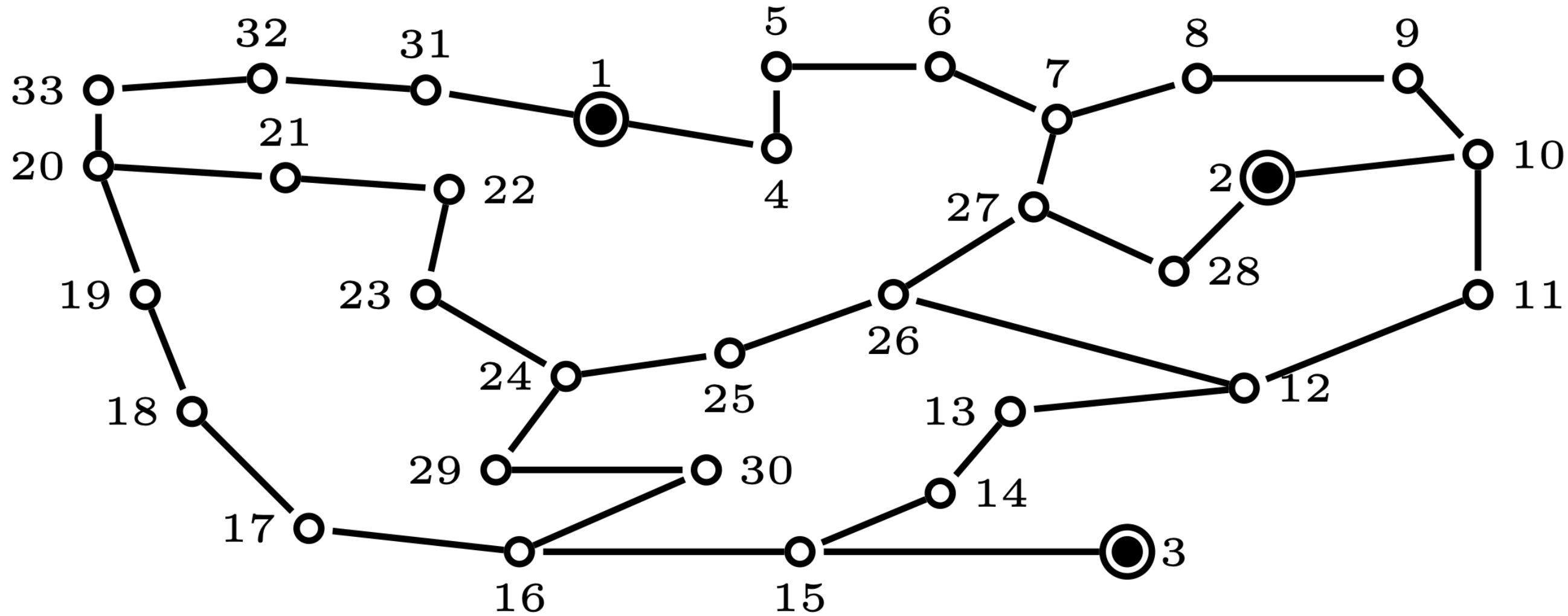} \\ 
       \small (a) Original network. \\
       \\
       \includegraphics[width=0.9\linewidth,height=0.45\linewidth]{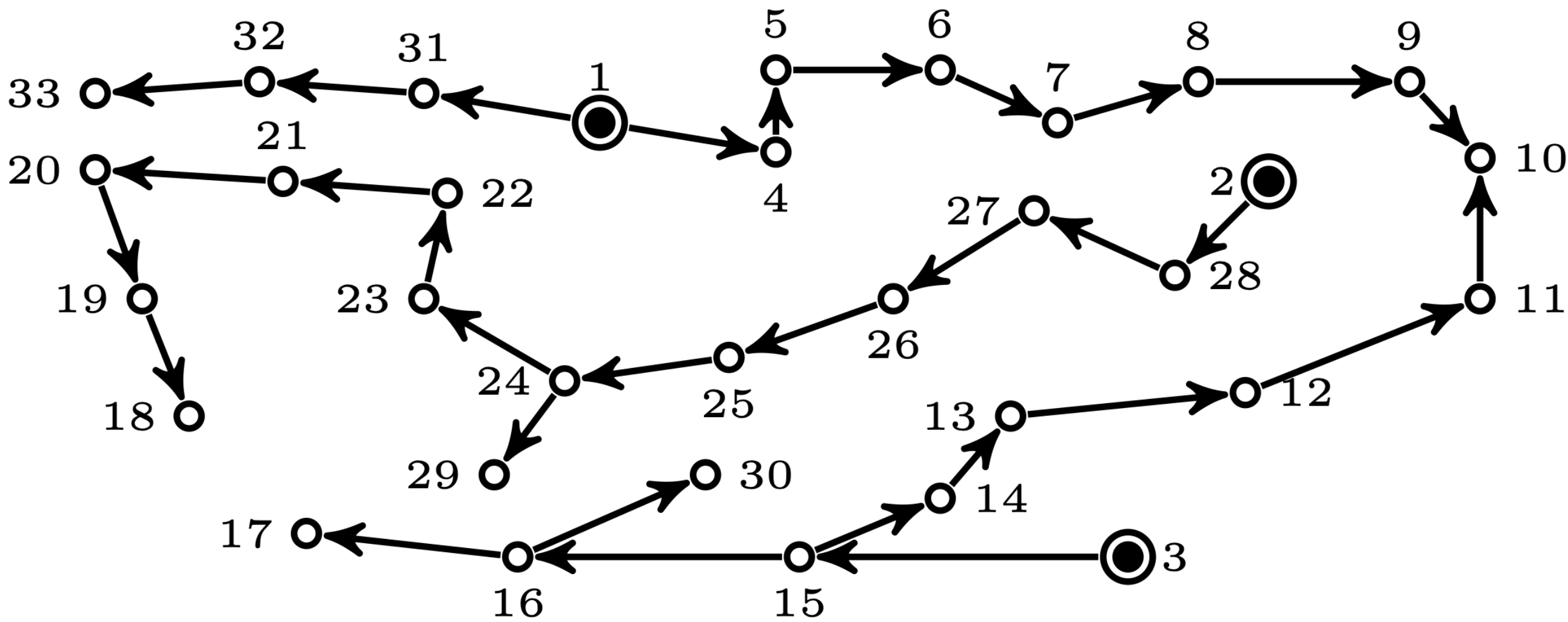} \\
       \small (b) Radial configuration.
    \end{tabular}
    \medskip
    \caption{{\small  In the optimal reconfiguration problem, the highlighted nodes in dark are the sources and the remaining nodes are the sinks. In radial configuration, some nodes, e.g., sink node 10 may receive receive input from two different edge; despite that there is no cycle in the graph. The network used here is the IEEE 33 network~\cite{ieee33}. }}  
    \label{fig::network_example} 
\end{figure}

We consider a network-flow distribution problem over a bidirectional distribution network $\mathcal{G}_D=\mathcal{G}(\mathcal{V}_D, \mathcal{E}_D)$ with $ |\mathcal{V}_D| = N $ nodes, $|\mathcal{E}_D| = m$ edges, a set of $n_g$ source nodes $\mathcal{V}_g \subset \mathcal{V}_D$, and $n_c = N - n_g $ sink nodes $\mathcal{V}_c = \mathcal{V}_D \backslash \mathcal{V}_g$, each with specified input and output.  We let $\vect{d}\in\real_{\geq0}^N$ be the output vector and $\vect{g}\in\real_{\geq 0}^N$ be the input vector, where $g_i\in\real_{>0}$ (resp. $d_j\in\real_{>0}$) for $i\in\mathcal{V}_g$ (resp. $j\in\mathcal{V}_c$) and $ g_i=0$ (resp. $d_j=0$) otherwise. The assumption is that inputs match outputs, i.e., $\sum_{i\in\mathcal{V}_g} g_i=\sum_{i\in\mathcal{V}_c} d_i$. The goal is to find an (oriented) \emph{radial configuration} that delivers the input flow from the source nodes to the sink nodes with minimal overall cost, see Fig.~\ref{fig::network_example}. This cost results from the `resistance' or `toll' along the edges, characterized as a quadratic function of the flow across them.

\begin{defn}[Set of radial configurations]
    A radial configuration is a polyforest\footnote{A polyforest (or directed forest or oriented forest) is a directed acyclic graph whose underlying undirected graph is a forest. A forest is a type of graph that contains no loops. Consequently, forests consist solely of trees that might be disconnected, leading to the term `forest' being used \cite{forests}.} that includes all the nodes $\mathcal{V}_D$, has roots at $\mathcal{V}_g$ and the undirected version of its edges are subset of $\mathcal{E}_D$. We denote the set of these polyforest digraphs by $\mathcal{F}(\mathcal{G}_D,\mathcal{V}_g)$. For brevity, when clear from context, we will use only $\mathcal{F}$. \boxend
\end{defn}\label{def::radial}

\smallskip
The problem of interest can be formalized as 
\vspace{-0.08in}
\begin{subequations}\label{eqn::problem1}
\begin{align}
&\min ~\sum\nolimits_{(i,j)\in\mathcal{S}} C_{i,j}\cdot x_{i,j}^2,\quad \text{subject to}\\
     &~~ \mathcal{G}(\mathcal{V}_D,\mathcal{S})\in\mathcal{F},\quad \text{and}\quad A(\mathcal{S})\, \vect{x} ={\vect{g}} - {\vect{d}} \label{eqn::problem1-kirchof}
\end{align}
\end{subequations}
where $x_{i,j}\in\real_{>0}$ is the flow across the link $(i,j)$, $A(\mathcal{S})$ is the incidence matrix of the radial configuration (a decision variable of the optimization problem), $C_{i,j}$ is the coefficient of the cost of edge $(i,j)\in\mathcal{S}$. The constraints confine the radial configuration to $\mathcal{F}$ and enforce the flow conservation (Kirchhoff's law) at the nodes. Throughout this paper, we assume that there are several radial configurations for which the input can meet the specified output. Thus, the feasible solution set of the optimization problem~\eqref{eqn::problem1} is non-empty.

\setlength{\textfloatsep}{3pt}

Problem~\eqref{eqn::problem1} is highly relevant to various potential flow applications, including natural gas, water, and electricity distribution networks. For instance, in power systems, the efficient and reliable distribution of energy has become increasingly critical with the growing integration of renewable energy and distributed generation sources. Modern power distribution networks, comprising multiple distributed generators, must operate in a radial configuration to adhere to engineering and safety standards. Minimizing energy loss within these networks is vital for both economic viability and environmental sustainability. Consequently, radial configurations cannot be chosen arbitrarily; they must be designed to ensure feasibility and to optimize energy loss (cost of operation).

Optimal radial reconfiguration problems are NP-hard problems, primarily due to the exponential growth in possible configurations as the number of distribution links increases. Specific instance (1) is often cast as Mixed-Integer Non-Linear Programming (MINLP) because of problem-specific side constraints, but for the purposes of this paper we focus on this linear abstraction to address the salient complexities associated with modeling \eqref{eqn::problem1-kirchof}.
These problems remain computationally intensive, often requiring extensive computational time and lacking guarantees of optimality \cite{Dong2021}.

To address these challenges, this paper constructs a polynomial-time 
solution for~\eqref{eqn::problem1} by leveraging the graph-like characteristics of distribution systems and how potential flows naturally navigate through a network. Our algorithm uses a greedy radial construction process starting from source nodes, incrementally adding edges while considering the flow across constructed components and the demanded output of the remaining nodes to reach. We draw inspiration from the similarities between electric flow in power networks and random walks~\cite{randomwalk}, developing a novel `sampling' method\footnote{Although our radial configuration construction is deterministic, we use `sampling' as a conceptual analogy.} for constructing radial configurations. In the random walk approach to describing electricity distribution in a given network, a weight proportional to the inverse of the resistance along the corresponding link in the power network is assigned to each link. This creates a notion of the edge weights being the conductance of the edge. Just as electricity flows through paths of least resistance, a random walk probabilistically selects paths based on transition probabilities influenced by edge weights. 

Our method assigns weights to edges based on cost and downstream output, ensuring flow distribution aligns with network requirements and effectively delivers a radial configuration that meets demand. This methodology ensures the process aligns with both the physical properties of flow and the optimization constraints of the problem, enabling the efficient identification of feasible and near-optimal radial configurations in polynomial time. Our particular innovation is incorporating a mechanism (\textsf{Net-Concad} function explained in Section~\ref{sec:net-concad}) in our incremental `sampling' process that addresses the shortsightedness of greedy selections by informing the process of the comprehensive demand of the remaining nodes. Numerical examples demonstrate the effectiveness of our proposed algorithm.

\begin{figure}[t]
    \centering
    \begin{tabular}{c c c}
    \includegraphics[width=0.25\linewidth]{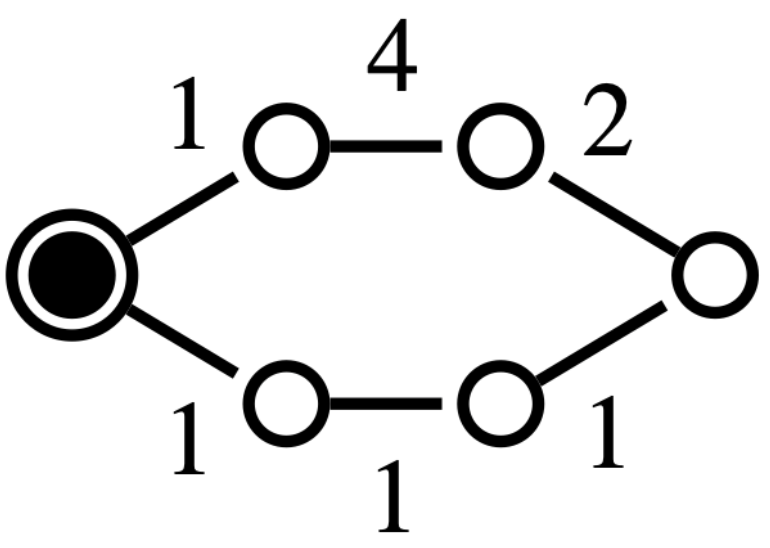} & \includegraphics[width=0.25\linewidth]{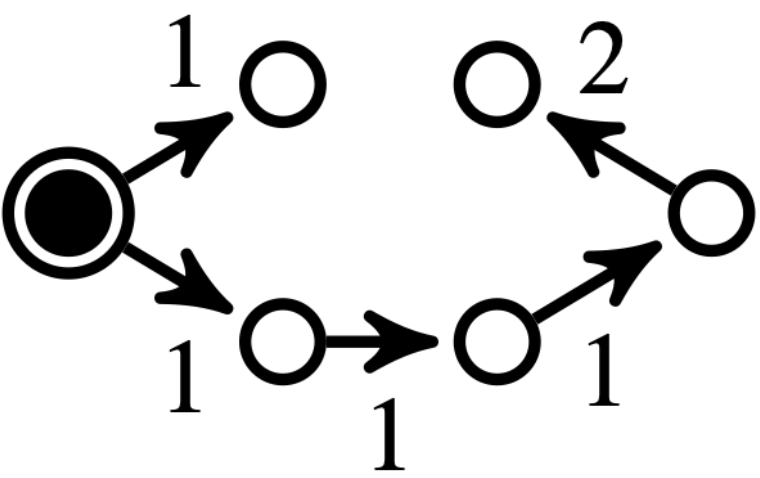} & \includegraphics[width=0.25\linewidth]{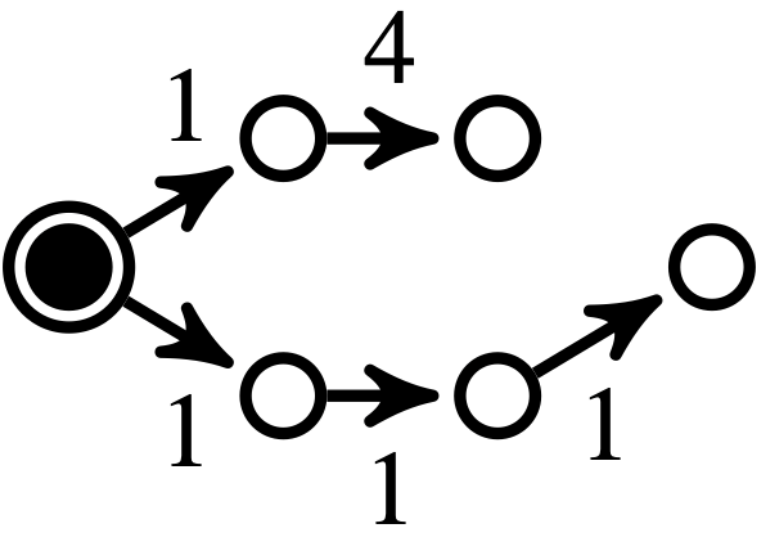} \\
         \small (a) $\mathcal{G}_D$ & \small (b) MST & \small (c) $\mathcal{G}(\mathcal{V}_D,\mathcal{S}^\star)$
    \end{tabular}
    \medskip
    \caption{{\small Example where radial distribution constructed from MST (plot (b)) is not the minimum radial configuration, $\mathcal{G}(\mathcal{V}_D,\mathcal{S}^\star)$, (plot(c)). This phenomenon is due to the quadratic nature of the cost; let the demand at each consumer be $d$ and the generator node, highlighted in bold, can supply input $5d$, in network (b) the cost is $1\cdot(d)^2 + 5\cdot(4d)^2=81d^2$, meanwhile in network (c) the cost is $5\cdot(2d)^2 + 3\cdot(3d)^2=47d^2$.}}
    \label{fig::quadratic_effect}
\end{figure}


\begin{figure}[t]
    \centering
    \begin{tabular}{c c c}
    \includegraphics[width=0.25\linewidth]{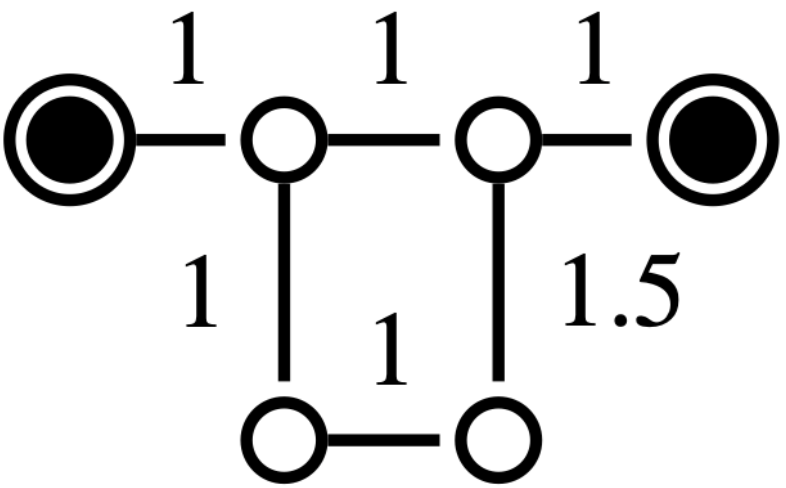} & \includegraphics[width=0.27\linewidth]{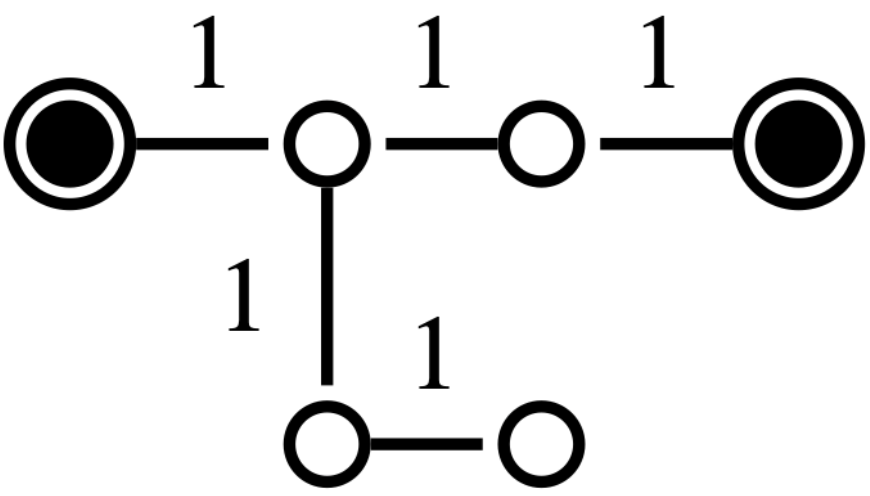} & \includegraphics[width=0.27\linewidth]{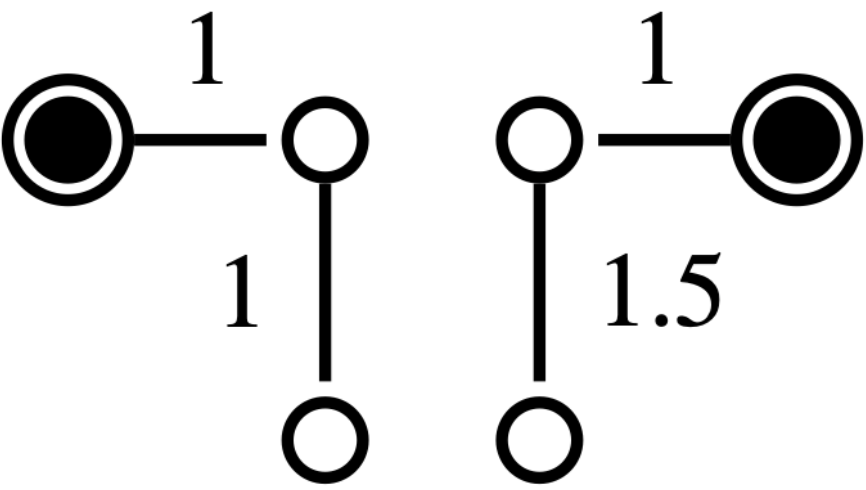} \\
         \small (a) $\mathcal{G}$ & \small (b) MST & \small (c) MSF
    \end{tabular}
    \medskip
    \caption{{\small Example where MSF results is a better outcome than MST. }}
    \label{fig::complexity_msf}
\end{figure}

\emph{Related work}: Several methods in the literature, especially in power networks applications, leverage topological properties to address the reconfiguration problem \cite{clark}. For example, \cite{khodabakhsh} \cite{spectral} use spectral clustering followed by local greedy search to identify radial configurations. In \cite{cuts}, the reconfiguration problem is linked to the maximum flow problem, based on Ford-Fulkerson’s work \cite{frodfulkerson}, which has been extensively studied in both single- and multi-source contexts \cite{maxflow}, \cite{maxflow_multi}, \cite{maxflow_multi2}. These approaches often include a repair procedure for unfeasible solutions, which can be inefficient. While some methods address minimum-cost distribution, radiality remains a critical constraint \cite{non_radial}, and cycle-breaking methods used to tackle it offer limited guarantees for large graphs \cite{cycle}. Other approaches, such as those based on the minimum spanning tree (MST) problem, face challenges due to the constraints of the reconfiguration problem, making the MST approach NP-hard in this context (see Fig.~\ref{fig::quadratic_effect}). Moreover, although algorithms like Kruskal's or Prim's \cite{tate2016prim} can find MSTs in polynomial time, they lose their optimality when multiple sources are involved, note that the minimum spanning forest (MSF) is not necessarily a subset of the MST (see Fig.~\ref{fig::complexity_msf}).

\setlength{\textfloatsep}{3pt}
\begin{figure*}[t!]
    \centering  
    \begin{tabular}{p{5cm} p{5cm} p{4.5cm}}
    \begin{minipage}{0.25\textwidth}
        \centering 
        \includegraphics[width=\linewidth]{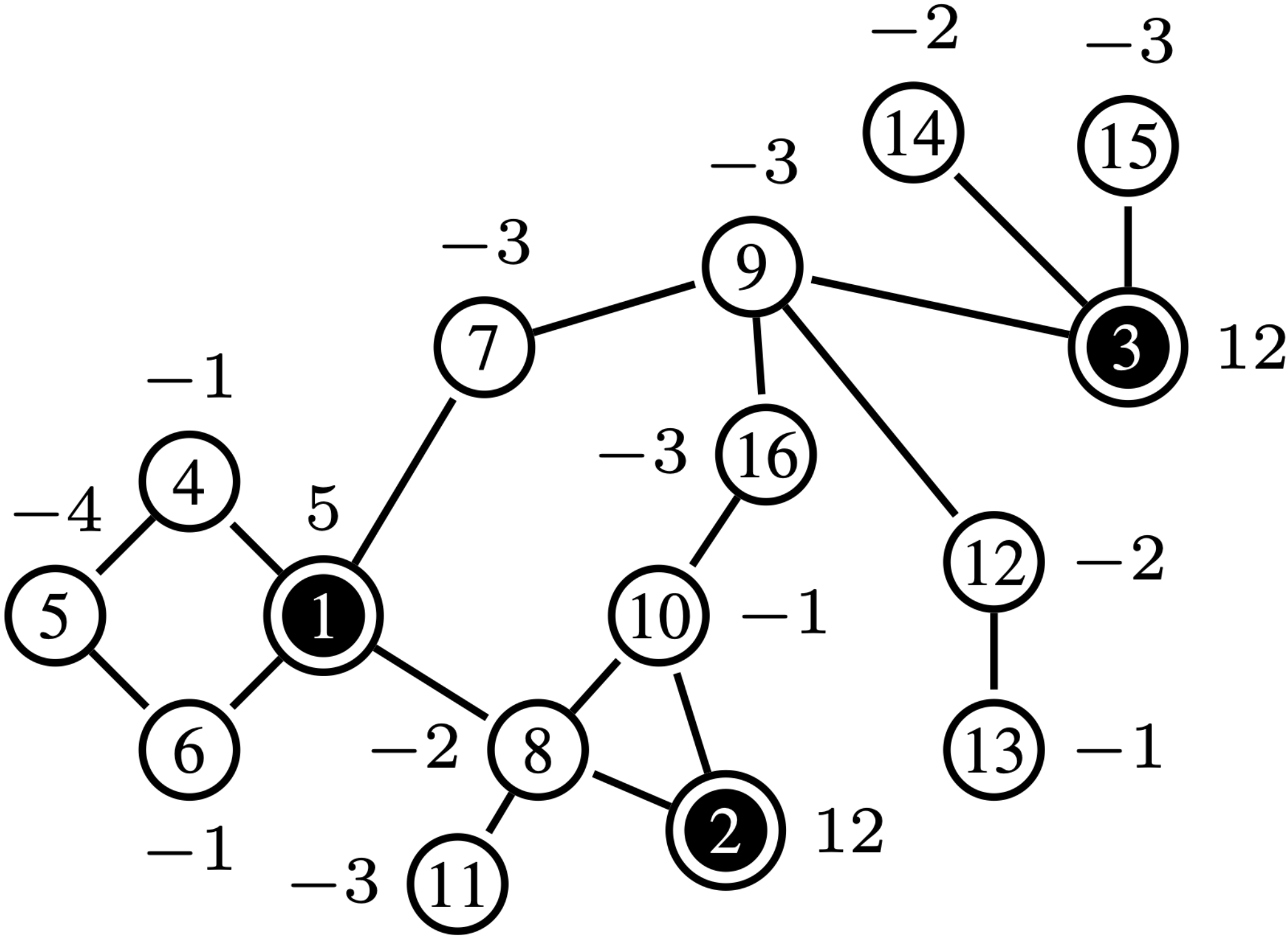} 
        \label{fig:compact0}
    \end{minipage} &
    \begin{minipage}{0.2\textwidth}
        \centering 
        \includegraphics[width=\linewidth]{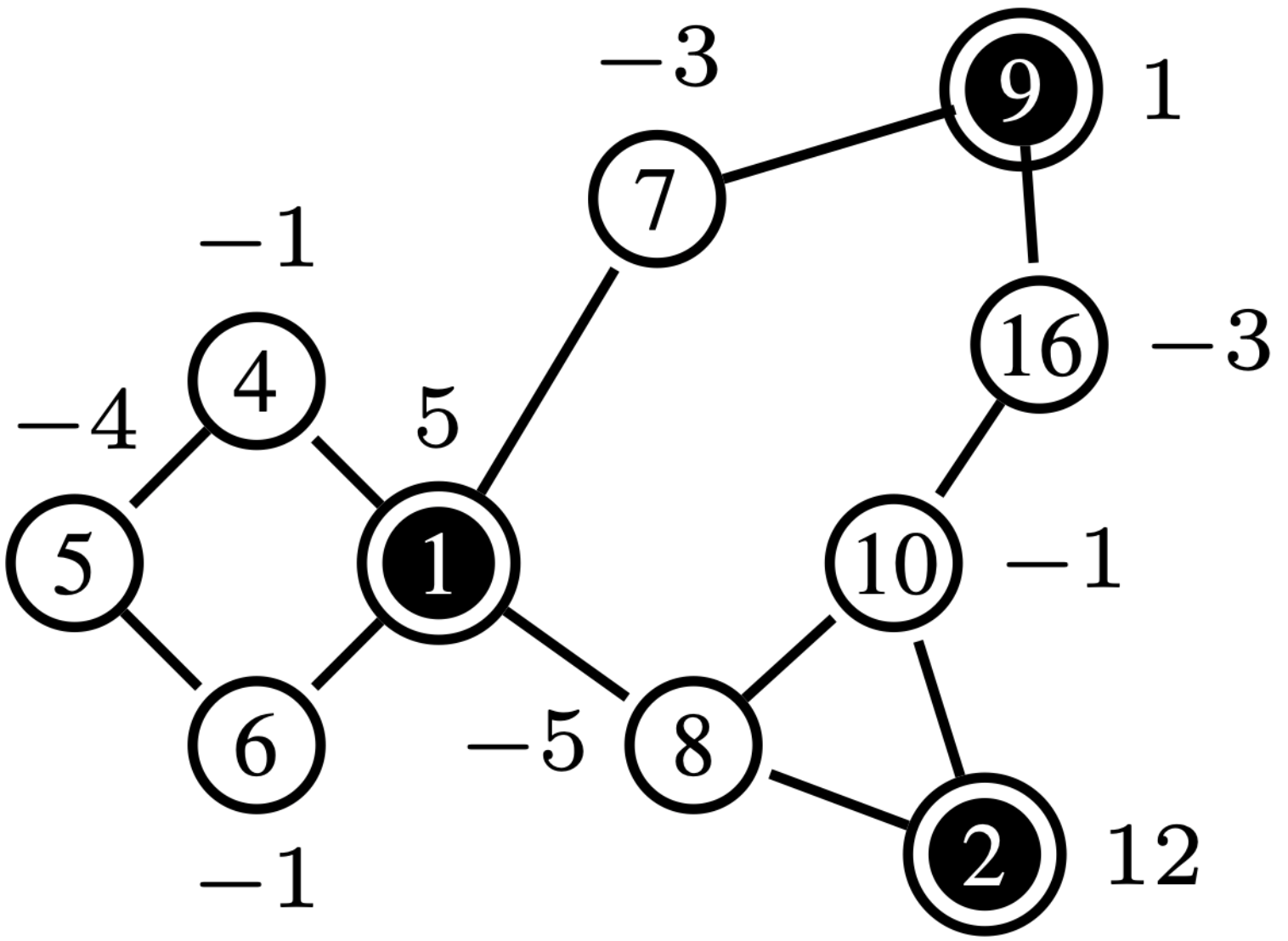} 
        \label{fig:compact1}
    \end{minipage} & 
    \begin{minipage}{0.25\textwidth}
        \centering 
        \includegraphics[width=\linewidth]{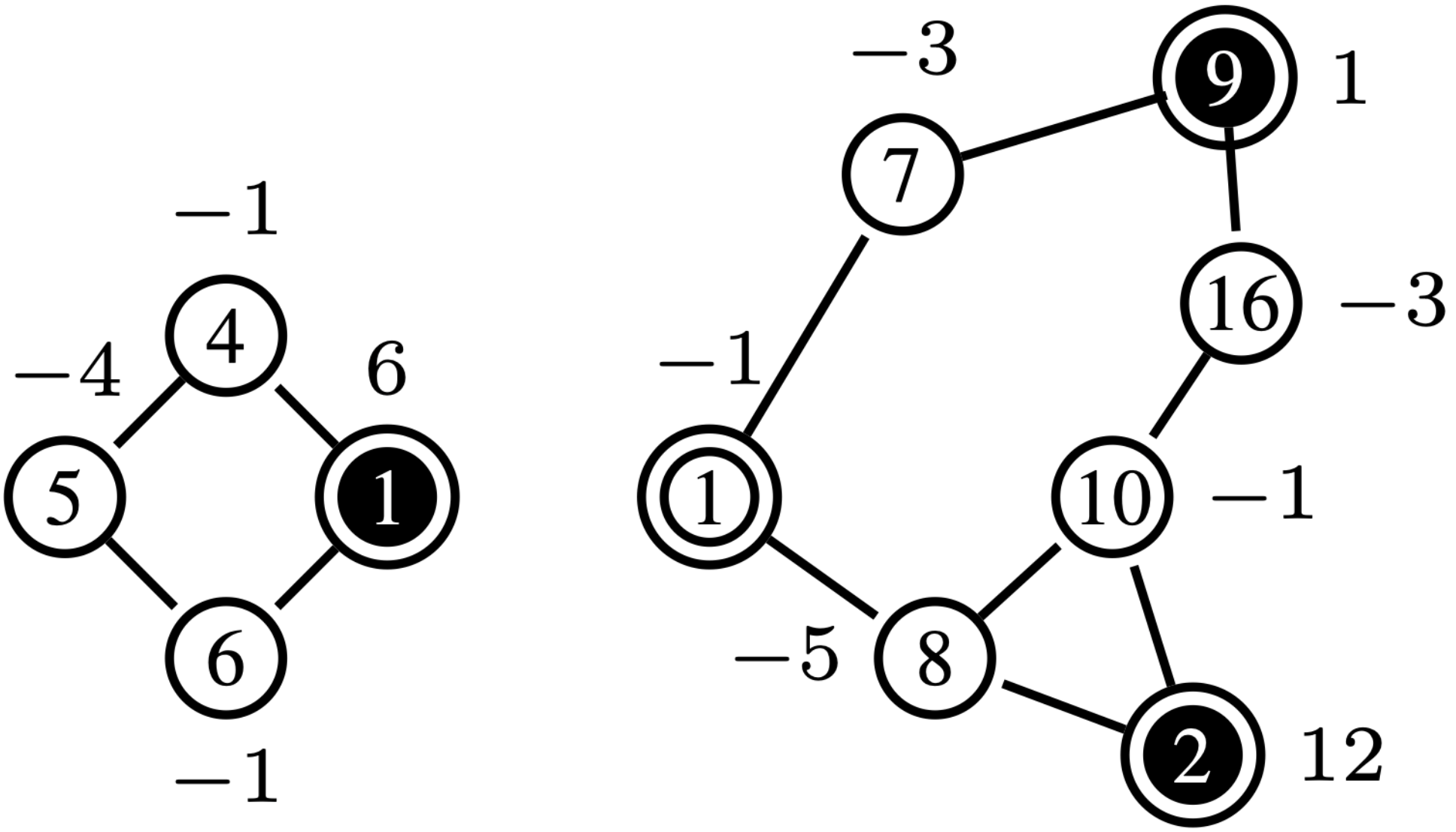} 
        \label{fig:compact2}
    \end{minipage}\\
    \small (a) Original graph $\mathcal{G}_0$. & 
    \small (b) \textsf{Pre-processor} output $\mathcal{G}_P$. & 
    \small (c) \textsf{Islander} splits $\mathcal{G}_P$, creating $\mathcal{G}^1$ and~$\mathcal{G}^2$. \\
    \centering
    \begin{minipage}{0.27\textwidth}
        \centering 
        \includegraphics[width=\linewidth]{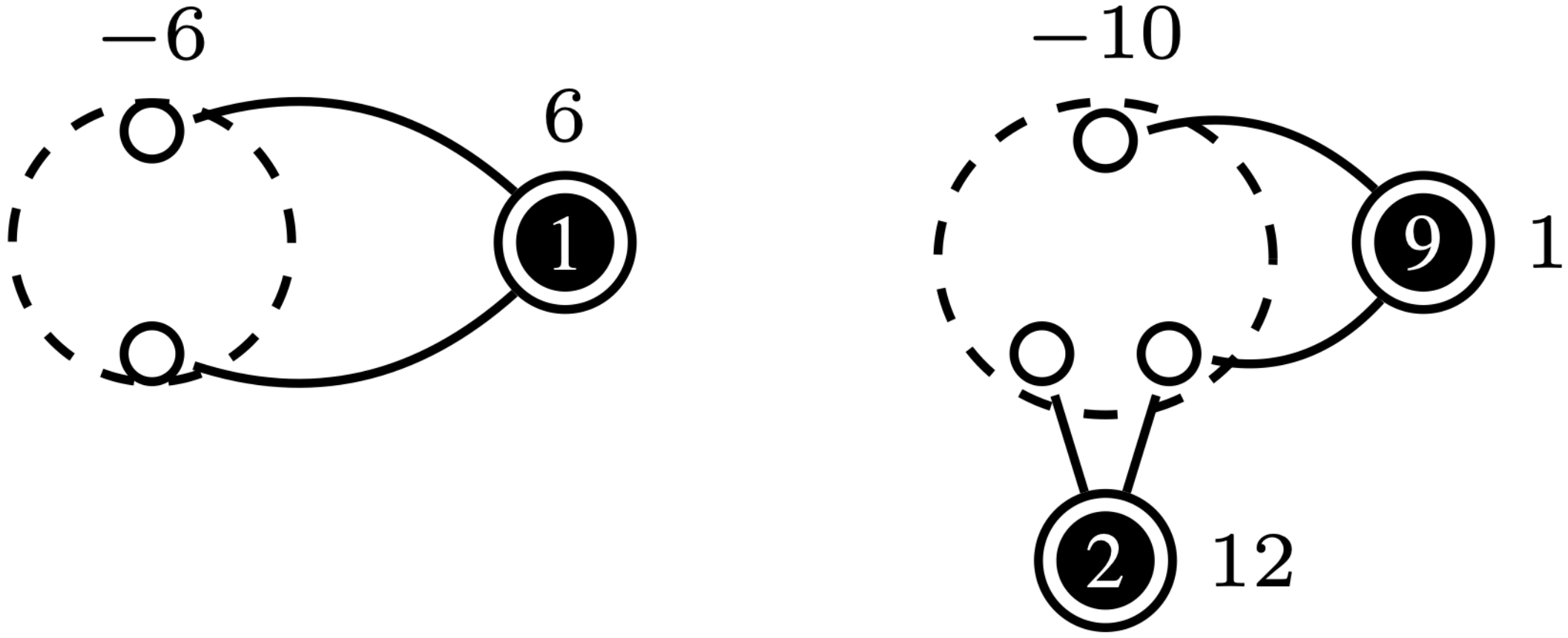} 
        \label{fig:compact3}
    \end{minipage} & 
    \begin{minipage}{0.18\textwidth}
        \centering 
        \includegraphics[width=\linewidth]{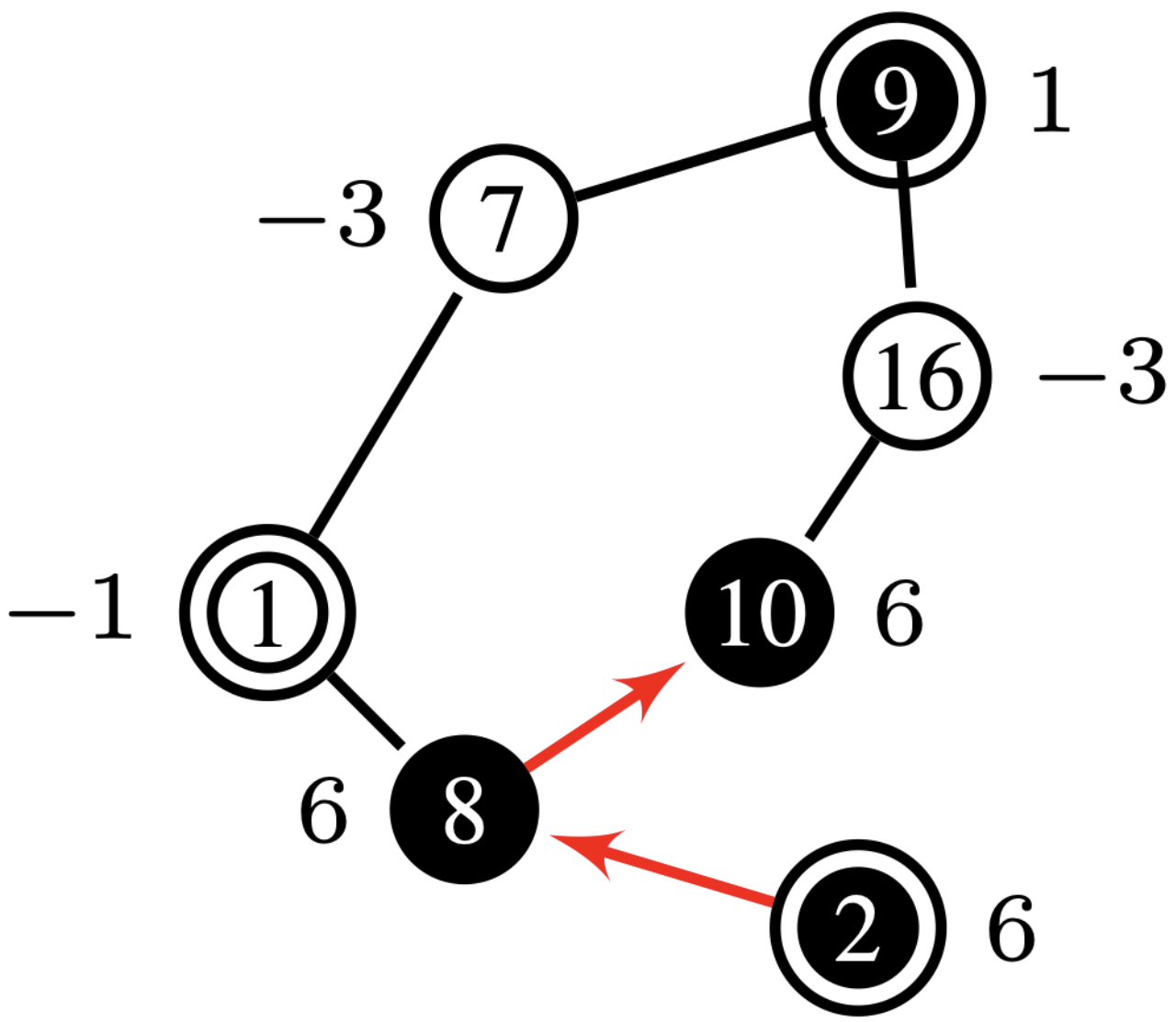} 
        \label{fig:compact4}
    \end{minipage} &
    \begin{minipage}{0.18\textwidth}
        \centering 
        \includegraphics[width=\linewidth]{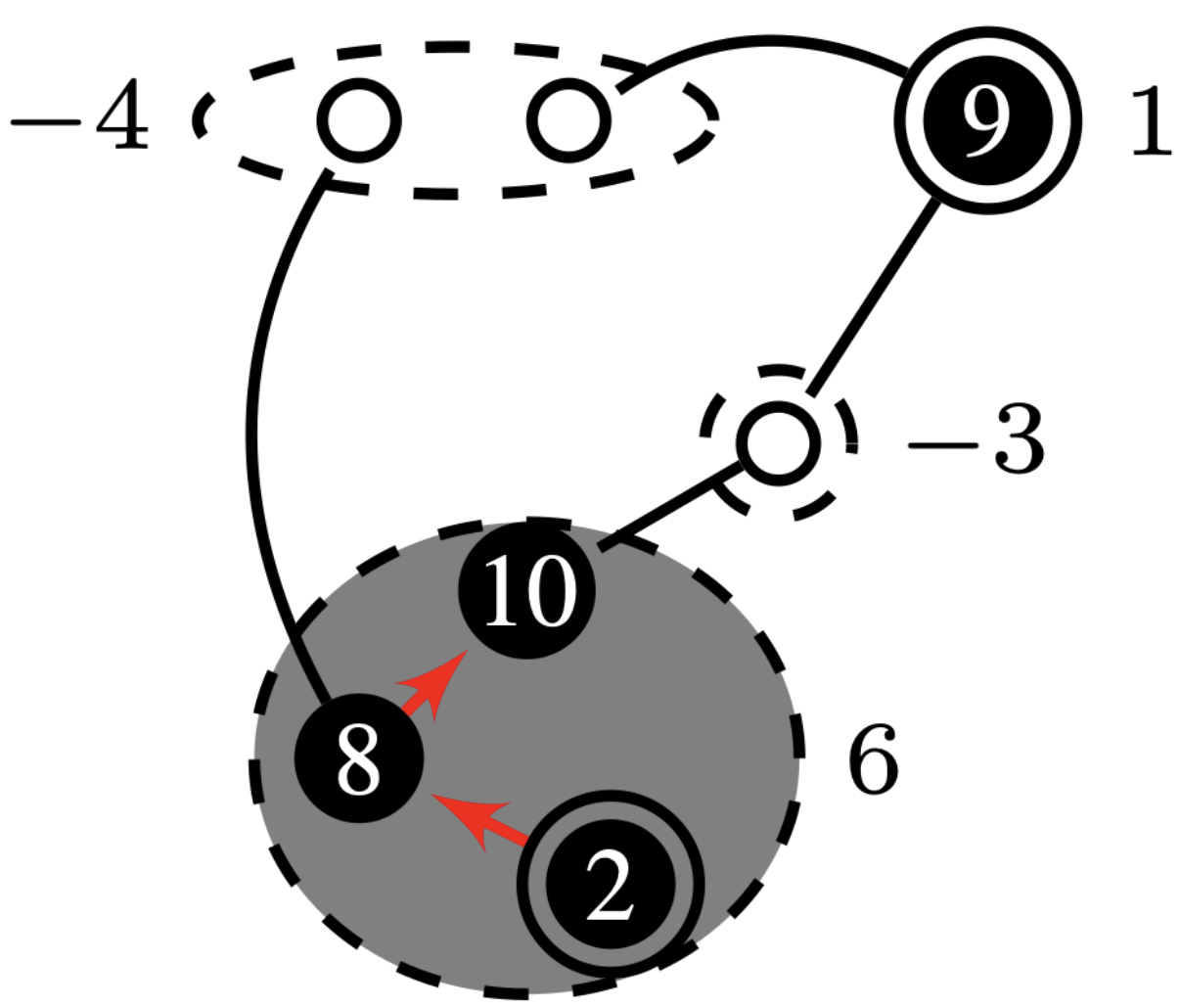} 
        \label{fig:compact5}
    \end{minipage}  \\
    \small (d) \textsf{Net-Concad} creates $\bar{\mathcal{G}}^1,\bar{\mathcal{G}}^2$ for the sub-graphs given in plot (b). & 
    \small (e)  $\mathcal{G}^2$ after multiple sampling. & 
    \small (f) \textsf{Net-Concad} output when its input graph in plot (e).
    \end{tabular}
    \medskip
    \caption{{\small Demonstration of how \textsf{Pre-processor},\textsf{Islander} and \textsf{Net-Concad} function. Filled solid nodes represent sources while others represent the sinks. The number next to the nodes show corresponding $p_i$. In plot (b) $p_9$ and $p_8$ are adjusted to reflect, respectfully, excess input to deliver through node 9 to $\mathcal{G}_p$ and extra demand at node $8$ to supply to the removed pendent node $11$. In plot (c), after partitioning the graph, node $1$ assumes different roles in each sub-graph. In plot (d) and (f) dashed circles represent super nodes. In plot (e), $\mathcal{T}_1=\mathcal{G}(\{9\},\{\})$ and $\mathcal{T}_2=\mathcal{G}(\{2,8,10\},(2\to8),(8\to10))$; note that here, \textsf{Sampler} has removed edge $(2,10)$ to avoid cycle. 
    }}    
    \label{fig:connected_comps}
\end{figure*}

\section{Algorithm design}
\label{sec::algorithm}

Our proposed algorithm follows a greedy structure, typical of MST algorithms. Starting from an empty edge set $\mathcal{S}$, we iteratively add edges connecting the nodes, guided by a heuristic to ensure feasibility. Inspired by the similarity between flow in distribution networks and random walks \cite{randomwalk}, we use a dual graph where edge weights represent the inverse of distribution costs. Our random walk starts at source nodes, assigning weights (intuitively interpreted as probabilities) to edges and adding the one with the highest probability of traversal to the growing forest.

We introduce some preliminary steps to sampling to compensate shortsightedness in greedy decision making algorithms by reducing the set of possible edges to select at each iteration. The first step is implementing the \textsf{Pre-processor} function which is followed by the \textsf{Islander} function to simplify the search space. Then, we apply  sequentially \textsf{Net-Concad} and \textsf{Sampler} functions for finite number of times to construct the radial configuration. These four functions constitute the elements of our proposed algorithm, which we call \textsf{FORWARD} as defined in the title of the paper. These functions are explained one by one below. Before that we introduce some notations. We define the input vector $\vect{p}$, associated with the nodes $\mathcal{V}_D$. This vector will be updated at each iteration of the algorithm, thus we denote it by $\vect{p}^t$ at each iteration $t$. We initialize at $p_i^0=-d_i<0$ for $i\in\mathcal{V}_c$ and $p_i^0=g_i>0$ if $i\in\mathcal{V}_g$. 
In what follows, we define the operator $\mathcal{V}(\mathcal{S})$ as the nodes of the sampled edge set $\mathcal{S}$, $\mathcal{E}(\mathcal{V}_i)$ as the set of edges in $\mathcal{E}_D$ interconnecting nodes in $\mathcal{V}_i$ and $\mathcal{N}(\mathcal{V}_i)$ as the set of nodes connected to nodes in $\mathcal{V}_i$. Also, $(i\to j)$ denotes a directed edge indicating  flow from node $i$ to $j$. Given a directed edge set $\mathcal{S}$, $\mathcal{U}(\mathcal{S})$ returns the undirected edge version of $\mathcal{S}$.

\subsection{Pre-processor}

Distribution networks often exhibit a small-world graph structure \cite{Hartmann2021}, which implies the presence of numerous nodes with low degree of connectivity. Consequently, this structure suggests that some radial sub-graphs are likely part of the original network $\mathcal{G}_0=\mathcal{G}_D$. It is important to note that the edges of any radial sub-graph within $\mathcal{G}_0$ conforms a unique solution, see Section~\ref{sec::feasibility}, so it is more efficient to aggregate them together. To address this, we introduce the \textsf{Pre-processor} function, which removes these trivially must-sampled components from the distribution network and adds them to the growing~$\mathcal{S}$. 

The \textsf{Pre-processor} function, described in Algorithm~\ref{alg::preprocessor}, takes as input a graph $\mathcal{G}_0$ and the associated input vector of the nodes, $\vect{p}^0$. Then, it samples every edge connecting single degree nodes, so-called pendant nodes (e.g., edges $(8,11)$, $(9,12)$, $(12,13)$, $(3,14)$, $(3,15)$ and $(9,3)$ in Fig.~\ref{fig:compact0}), and adds these edges to $\mathcal{S}$. Once these edges and nodes are added to the solution set $\mathcal{S}$, \textsf{Pre-processor} removes the pendant nodes and their connecting edges, redistributes their input/output back to the parent nodes, and repeats the process until no pendant nodes remain in the updated graph. The result of implementing the \textsf{Pre-processor} function on the network in Fig.~\ref{fig:compact0} is shown in Fig.~\ref{fig:compact1}. After the network is simplified by \textsf{Pre-processor}, all nodes in the resulting graph $\mathcal{G}_P$ are at least 2-connected.

\begin{algorithm}[t]
{\footnotesize
\caption{{\small\textsf{Pre-Processor}}}
\begin{algorithmic}[1]
\Require Bidirectional graph $\mathcal{G}_0=\mathcal{G}_D$, input vector $\vect{p}^0$
\State $\mathcal{S} \leftarrow \emptyset$
\State $\mathcal{E}_P \leftarrow \mathcal{E}_D$
\While{exists pendant nodes}
\State Pick a pendent edge $(i,j)\in\mathcal{E}_P$, let $i$ denote the pendent node
\If{$p_i\geq0$}
\State $\mathcal{S}_0\leftarrow \mathcal{S}_0\cup\{(i\rightarrow j)\}$ 
\Else
\State $\mathcal{S}_0\leftarrow\mathcal{S}_0\cup\{(j\rightarrow i)\}$ 
\EndIf
\State $p_j=p_j+p_i$
\State $\mathcal{E}_P\leftarrow \mathcal{E}_P\setminus\{(i,j)\}$
\EndWhile
\State $\mathcal{G}_P\leftarrow \mathcal{G}(\mathcal{V}(\mathcal{E}_P),\mathcal{E}_P)$
\State Reshape $\vect{p}$ to include only the elements corresponding to nodes of $\mathcal{G}_P$
\State \Return $\mathcal{S}_0$, $\mathcal{G}_P$, $\vect{p}$
\end{algorithmic}
\label{alg::preprocessor}
}
\end{algorithm}


\subsection{Islander}
\label{sec:islander}
\vspace{-1mm}
\textsf{Islander}, described in Algorithm~\ref{alg::islander}, takes as input graph $\mathcal{G}_P$ generated by \textsf{Pre-processor}, the source nodes in $\mathcal{G}_P$ and its input vector $\vect{p}$, and proceeds to partition the graph into disjoint sub-graphs induced by removing the articulation source nodes \footnote{An articulation node in a graph is a node which, if removed, graph is disconnected. A bi-connected graph is a graph with no articulation nodes.}. The results are $\mathcal{G}^\ell$,  $\ell\in\{1,\cdots,L\}$, where $L$ is the number of sub-graphs obtained in the process. The input vector $\vect{p}$ of each articulation super node will then be adjusted to balance the overall input-output in each sub-graph; the result of this process can change the role of the super nodes from a source to sink or vice versa. For example, after feeding the input graph in Fig.~\ref{fig:compact1} whose source nodes are
$1$, $2$ and $9$, to \textsf{Islander}, the function partitions the graph into two sub-graphs as shown Fig.~\ref{fig:compact2}, in which the role of node $1$ in each graph is different. 

\textsf{Islander} enables us to search for radial configuration in each sub-graph in parallel, speeding up the process. Note that if two sub-graphs are only connected through an articulation node, and if both sub-graphs present a radial configuration, the overall graph remains radial.

\begin{algorithm}[t]
{\footnotesize
\caption{{\small\textsf{Islander}}}
\begin{algorithmic}[1]
\Require A graph $\mathcal{G}_P$, source nodes $\mathcal{V}_g$, input vector $\vect{p}$
\State Find nodes in $\mathcal{V}_P\cap\mathcal{V}_g$ which are articulation nodes of $\mathcal{G}_P$
\State Partition $\mathcal{G}_P$ at articulation nodes creating $\mathcal{G}^\ell=\mathcal{G}(\mathcal{V}^\ell,\mathcal{E}^\ell)$, $\ell\in\{1,\cdots,L\}$ \State Define an input vector $\vect{p}^\ell$, which contains the elements of $\vect{p}$ that corresponds to nodes in $\mathcal{G}^\ell$, $\ell\in\{1,\cdots,L\}$.
\State Compute and update $p_i^\ell$ for each articulation point $i$ in each partition $\ell$
\State $\mathcal{V}_g^\ell=\{i\in\mathcal{V}^\ell|p_i^\ell>0\}$, ~~$\ell\in\{1,\cdots,L\}$
\State \Return $(\mathcal{G}^1,\mathcal{V}_g^1,\vect{p}^1),\cdots,(\mathcal{G}^L,\mathcal{V}_g^L,\vect{p}^L)$ 
\end{algorithmic}
\label{alg::islander}
}
\end{algorithm}

\subsection{Net-Concad}
\label{sec:net-concad}

\textsf{Net-Concad}, described in Algorithm~\ref{alg::netconcad}, takes as input a graph $\mathcal{G}^\ell_P$, the existing sampled radial polytrees $\mathcal{T}_1, \cdots, \mathcal{T}_b$ in $\mathcal{G}^\ell_P$ and its input vector $\vect{p}^\ell$.
The existing radial polytrees, starting from the source nodes, are incrementally built by the \textsf{Sampler} to channel the input flow from the sources to the sinks. Here, $b$ represents the number of polytrees already formed in the graph, which will be, at most, equal to the number of sources in the graph. An important notion we employ is that all nodes within a polytree have positive input to supply. Therefore, based on the flow vector $\vect{p}^t$, at iteration $t$ some nodes can assume ``pseudo" roles. See Fig.~\ref{fig:compact5} for an example of input graph and existing polytrees in it.

After partitioning the graph by \textsf{Islander}, \textsf{Net-Concad} proceeds to condensing each sub-graph $\bar{\mathcal{G}}^\ell$ by removing the formed polytrees (grouped in super source nodes) and turning the resulted connected components \cite{conncomp} into super sink nodes as shown in  Fig.~\ref{fig:compact3} and Fig.~\ref{fig:compact5}.
In this new configuration we can assume, without loss of generality, that each sub-graph is a bi-connected and quasi-bipartite graph in terms of super source nodes as one group and super sink nodes as the other.

\begin{defn}
    A quasi-bipartite graph is a graph where all its vertex can be divided into two disjoints subsets such as most of the edges exists between these subsets. However, conversely to bipartite graphs, few edges may occur within one of the subsets. 
\end{defn}

\begin{algorithm}[t]
{\footnotesize
\caption{{\small\textsf{Net-Concad}}}
\begin{algorithmic}[1]
\Require A graph $\mathcal{G}^\ell$
\State Define $\mathcal{V}_g\leftarrow\{i:p_i>0\}$ and $\mathcal{V}_c\leftarrow\{i:p_i\leq0\}$, $\forall i\in\mathcal{V}^\ell$
\State Define $\mathcal{G}_g\leftarrow\mathcal{G}(\mathcal{V}_g,\mathcal{E}(\mathcal{V}_g))$ and $\mathcal{G}_c\leftarrow\mathcal{G}(\mathcal{V}_c,\mathcal{E}(\mathcal{V}_c))$ 
\State $\bar{\mathcal{G}}_g,\bar{\mathcal{G}}_c \leftarrow$ Run \cite{conncomp} over $\mathcal{G}_g$ and $\mathcal{G}_c$
\State Re-connect sub-graphs $\bar{\mathcal{G}}_P\leftarrow\mathcal{G}(\bar{\mathcal{V}}_g\cup\bar{\mathcal{V}}_c, \mathcal{E}_D\setminus\{\mathcal{E}(\mathcal{V}_g\cup\mathcal{V}_c)\})$
\State \Return $\bar{\mathcal{G}}^\ell$
\end{algorithmic}
\label{alg::netconcad}
}
\end{algorithm}

It is important to note that each sub-graph $\mathcal{G}^{\ell}$ for $\ell \in \{1, \ldots, L\}$ generated by \textsf{Islander} is irreducible; specifically, the super source nodes introduced at each iteration of \textsf{Net-Concad} do not become articulation points. 
\begin{lem}
    \label{rem::irreducibility}
    Each partitioned sub-graph $\bar{\mathcal{G}}^\ell$ is an irreducible graph.
\end{lem}
\begin{proof}
    Since  each  $\bar{\mathcal{G}}^\ell$ is a 2-connected graph and growing the super source nodes does not affect this property, the graph does not include an articulation super source and thus it is irreducible. 
\end{proof}


\subsection{Sampler}

In order to choose which is the ``most probable" edge to sample from a set of candidates we proceed with calling \textsf{Sampler} function. The \textsf{Sampler} assumes that some weight $w_{i,j}$ is assigned to each candidate edge, which gets updated after sampling. These weights are defined by the problem context and the physical specificity of the flow. We give an example case in  Section~\ref{sec:numeric}, where the underlying problem is an electric power distribution.

In order to keep the growing polyforest in the feasible domain, \textsf{Sampler} needs to guarantee the available flow in the network does not get stuck in it. Therefore it operates by a priority queue $q$ which first samples edges connecting pendant sources in each partition. Then, it also prioritizes edges which connect nodes which can provide the output demand of its children, so that the balance $p_i^\ell+p_j^\ell\geq 0$. Finally, to prevent the creation of cycles, taking inspiration from Loop-erase Random-Walk algorithm \cite{Lawler1980}, we implement an 
\emph{edge-delete} procedure (line 4 in Algorithm~\ref{alg::sampler}) where, prior to sample any edge, the \textsf{Sampler} looks for edges in $\bar{\mathcal{E}}^\ell$, whose nodes are already included in one of the existing~$\mathcal{T}_i^\ell$.

\begin{algorithm}[t]
\footnotesize
\caption{\textsf{Sampler}}
\begin{algorithmic}[1]
\Require Sub-graph partition $\bar{\mathcal{G}}^\ell$, flow vector $\vect{p}^\ell$ and $\mathcal{S}^\ell$
\State Initiate priority queue $q\leftarrow\emptyset$
\State Compute weight $\forall e\in\mathcal{E}^\ell$ and add them to $q$
\State Sort $q$ in increasing order.
\State \emph{Edge-delete} procedure: delete all edges with both endpoints in $\mathcal{V}(\mathcal{S}^\ell)$
\State Push all edges $(i\rightarrow j)$ such that $p_i^\ell+p_j^\ell\geq 0$ to the top of $q$
\State Push all edges $(i\rightarrow j)$ such that $|\mathcal{N}(i)|=1$ to the top of $q$
\State $e^\star \leftarrow \textup{pop from } q$
\State \Return $e^\star$
\end{algorithmic}
\label{alg::sampler}
\end{algorithm}

\subsection{\textsf{FORWARD}}
After \textsf{Sampler} is done, \textsf{FORWARD}, presented in Algorithm~\ref{alg::algorithm}, proceed to update the information of the network. Note that the process is an adaptive procedure where the flow remaining in the system must decrease as sampling. Indeed, flow decreases proportionally to the output of the new node added to a polytree. Then \textsf{Net-Concad} and \textsf{Sampler} are applied again till convergence when all nodes are connected to a polytree, which will happen after $N-1$ iterations at~most.

\begin{rem}[Complexity of \textsf{FOWARD}]  Notice that \textsf{Pre-processor} has a complexity of $\mathcal{O}(n)$, \textsf{Net-Concad} runs in $\mathcal{O}(m+n)$ which obeys the complexity of the contraction algorithm, which the same complexity as \textsf{Islander}, and, finally, \textsf{Sampler} runs over the edges that belong to the neighbourhood of the growing forest which is, in the worst-case scenario, all the $m$ edges of $\mathcal{E}_D$, so that $\mathcal{O}(m)$. Therefore, proposed procedure has a complexity of $\mathcal{O}(n\cdot(m+n))$.  As distribution networks is a small-world graph, the number of edges is close to the number of nodes $m=c\cdot n$, therefore, complexity can be reduced to $\mathcal{O}(n^2)$ in the worst case scenario, which is polynomial time. However, it must be noticed that after \textsf{Pre-Processor} the remaining graph will have less than $n$ nodes and that \textsf{Islander} will partition the graph to even smaller sub-graphs which can be solved in parallel. Therefore, real implementation complexity can be significantly reduced. \boxend
\end{rem}

\begin{algorithm}[t]
{\footnotesize
\caption{FORWARD}
\begin{algorithmic}[1]
\Require Bidirectional graph $\mathcal{G}_0=\mathcal{G}_D$, input vector $\vect{p}^0$
\State $\mathcal{S} \leftarrow \emptyset$, 
\State $\mathcal{S}^0,\mathcal{G}_P,\vect{p} \leftarrow$ \textsf{Pre-Processor}($\mathcal{S},\mathcal{G}_0,\vect{p}_0$)
\State $(\mathcal{G}^1,\mathcal{V}_g^1,\vect{p}^1),\cdots,(\mathcal{G}^L,\mathcal{V}_g^L,\vect{p}^L)\leftarrow$\textsf{Islander}($\mathcal{G}_P,\vect{p}$)

\For{each partition $\ell\in\{1,\cdots,L\}$}
    \State $\mathcal{S}^\ell\leftarrow \emptyset$
    \State $\mathcal{T}^\ell_i = \mathcal{G}(\{i\},\{\})~~\forall i\in \mathcal{V}^\ell_g$
    \While{$|\mathcal{S}^\ell|\leq |\mathcal{V}^\ell|-1$}
        \State $\bar{\mathcal{G}}^\ell\leftarrow$\textsf{Net-Concad}($\mathcal{G}^\ell$)
        \State $e^\star\leftarrow$ \textsf{Sampler}($\bar{\mathcal{G}}^\ell,\vect{p}^\ell,\mathcal{S}^\ell$)
        \State $\mathcal{S}^\ell\leftarrow\mathcal{S}^\ell\cup\{e^\star\}$
        \State Update $\vect{p}^\ell$ and corresponding $\mathcal{T}^\ell_i$
    \EndWhile
\EndFor
\State $\mathcal{S}\leftarrow \bigcup_{\ell=0}^L\mathcal{S}^\ell$ 
\State \Return Radial configuration $\mathcal{G}(\mathcal{V}(\mathcal{S}),\mathcal{S})$
\end{algorithmic}
\label{alg::algorithm}
}
\end{algorithm}


\section{Feasibility analysis}
\label{sec::feasibility}

To find a feasible radial distribution configuration from $\mathcal{F}$ that can channel the inputs from the sources to the sinks and meet their demanded output is a non-trivial task. In what follows, we show that the \textsf{FORWARD} algorithm is guaranteed to generate such a feasible radial distribution configuration because of the inherent structure of the condensed dual graph. We start our demonstration with some auxiliary~lemmas and remarks.

\begin{lem}
   Any tree graph (undirected) admits a unique flow distribution from its sources to its sinks nodes as long as the total input meets the total demand, i.e., the (directed) polytree generated to deliver the input from the sources to the sinks is unique. 
    \label{lem::uniqueness}
\end{lem}
\begin{proof}
    For a connected acyclic graph with $n$ nodes, such as a tree graph, the incident matrix is full column rank $(n-1)$. Therefore, the null space of the incident matrix is trivial and thus there is a unique flow vector that satisfies the flow conservation constraint matrix equation.    
\end{proof}
Lemma~\ref{lem::uniqueness} asserts that in the sub-graphs separated from $\mathcal{G}_D$ by the \textsf{Pre-Processor} the (oriented) radial configuration to deliver the inputs to the sinks exists and is unique. Therefore, polytree grown by \textsf{Pre-Processor} belongs to all the configurations that are in the feasible set of optimization problem~\eqref{eqn::problem1}, including the optimal configuration.

 Note that the \textsf{Islander} function will not affect the sampling procedure as it re-distributes input vector $\vect{p}$ without disturbing the flow balance. In a similar way, applying \textsf{Net-Concad} to a graph $\mathcal{G}^\ell$ will not modify the sampling space. By construction, edges on the dual condensed graph, created by \textsf{Net-Concad}, $\bar{\mathcal{E}}^\ell$ are all the existing edges in $\mathcal{E}_P$. 
 Consequently, feasibility is guaranteed if we demonstrate that the \textsf{FORWARD} can always construct a feasible radial configuration in each $\mathcal{G}^\ell$  that delivers inputs in the sub-graph to the sinks (i.e., inputs meet the demands at the sinks).

The first result below shows that any digraph created by \textsf{Sampler} function through iterations is a polyforest (a radial configuration without a cycle).

\begin{lem}
   Any $\mathcal{G}(\mathcal{V}(\mathcal{S}^\ell),\mathcal{S}^\ell)$, $\ell\in\{1,\cdots,L\}$ generated by \textsf{FORWARD} is always a radial configuration. 
\end{lem}
\begin{proof}
The proof hinges on the fact that each edge is sampled at most once by \textsf{Sampler}, and the \emph{edge-delete} procedure within \textsf{Sampler} prevents the formation of cycles.
\end{proof}

The following lemma shows that any radial digraph created by \textsf{FORWARD} will visit every node in the graph.

\begin{figure}[t]
    \centering
    \begin{tabular}{c c c}
        \includegraphics[width=0.27\linewidth]{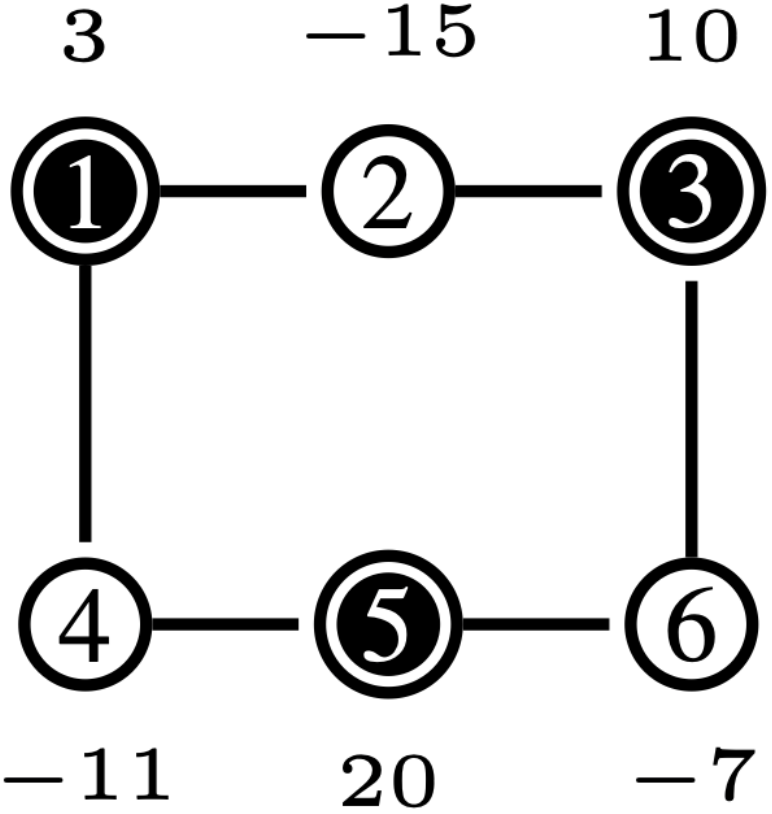} & \includegraphics[width=0.27\linewidth]{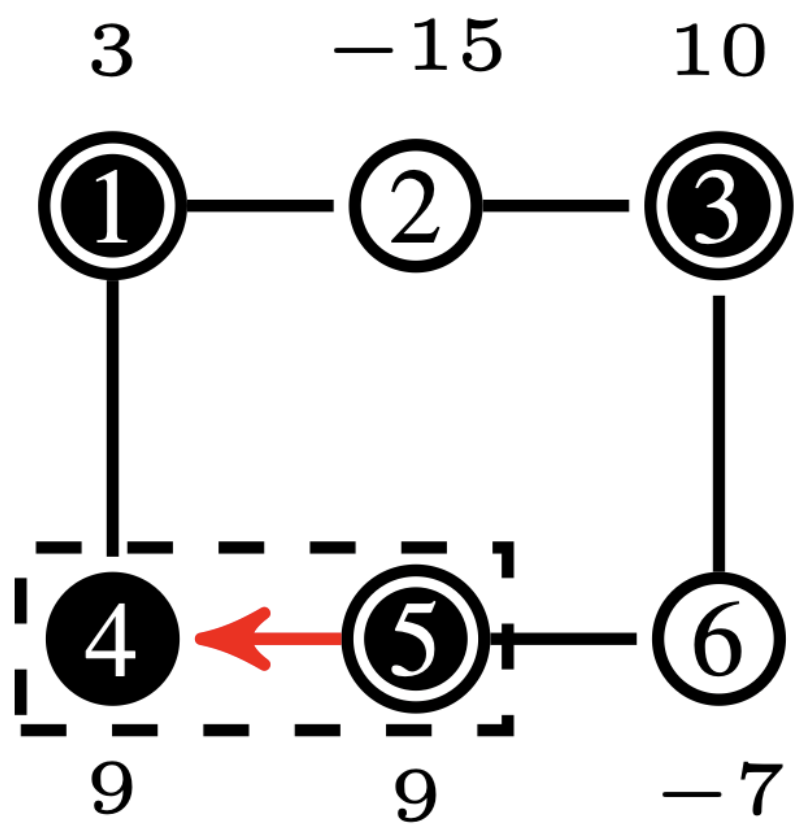} & \includegraphics[width=0.27\linewidth]{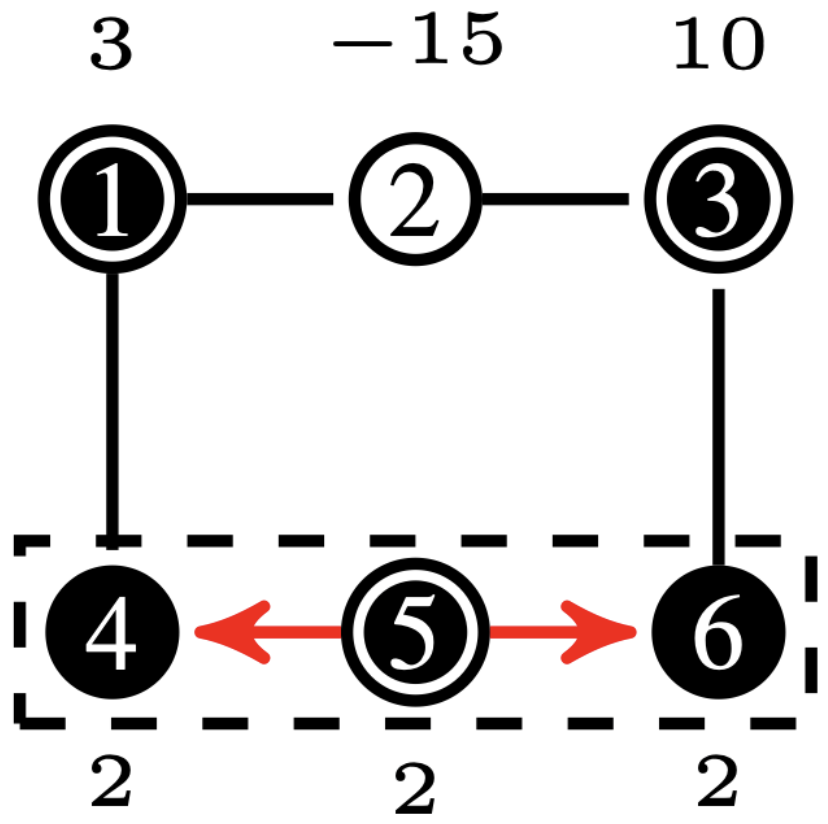} \\
        \small (a) ${\mathcal{G}}^\ell$ & \small (b) $1^{\textup{st}}$ step & \small (c) $2^{\textup{nd}}$ step  \\
        \includegraphics[width=0.27\linewidth]{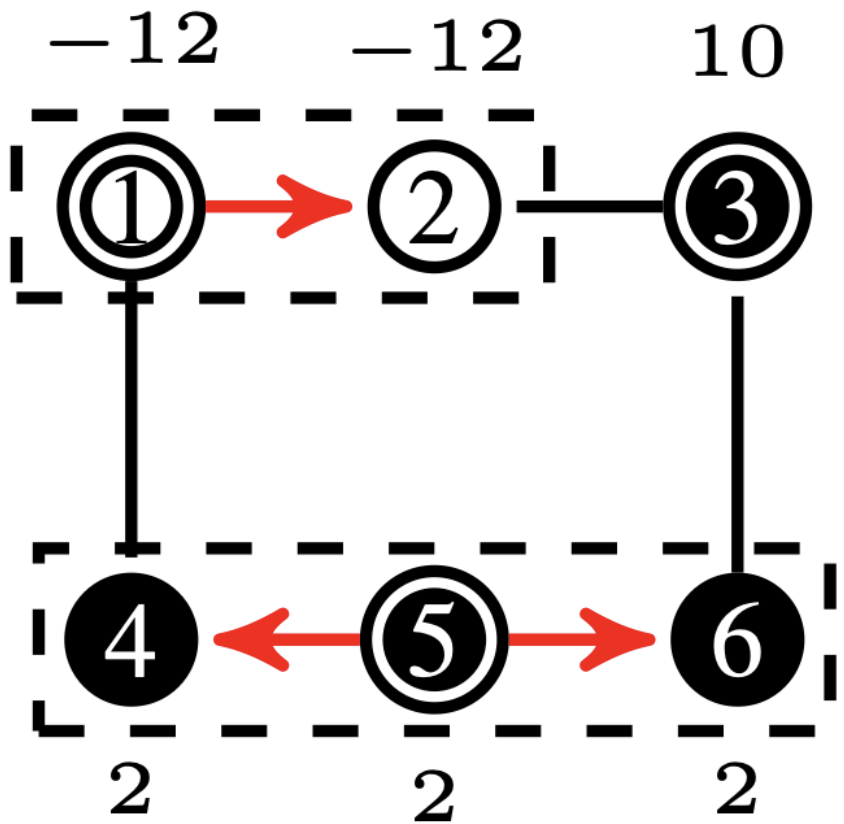} & \includegraphics[width=0.27\linewidth]{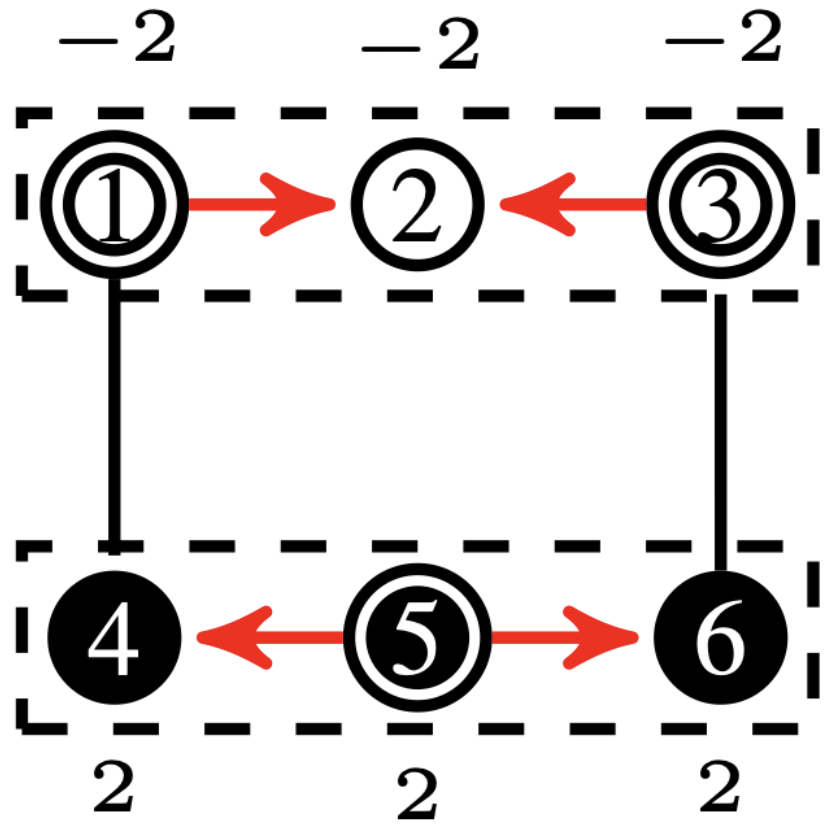} & \includegraphics[width=0.27\linewidth]{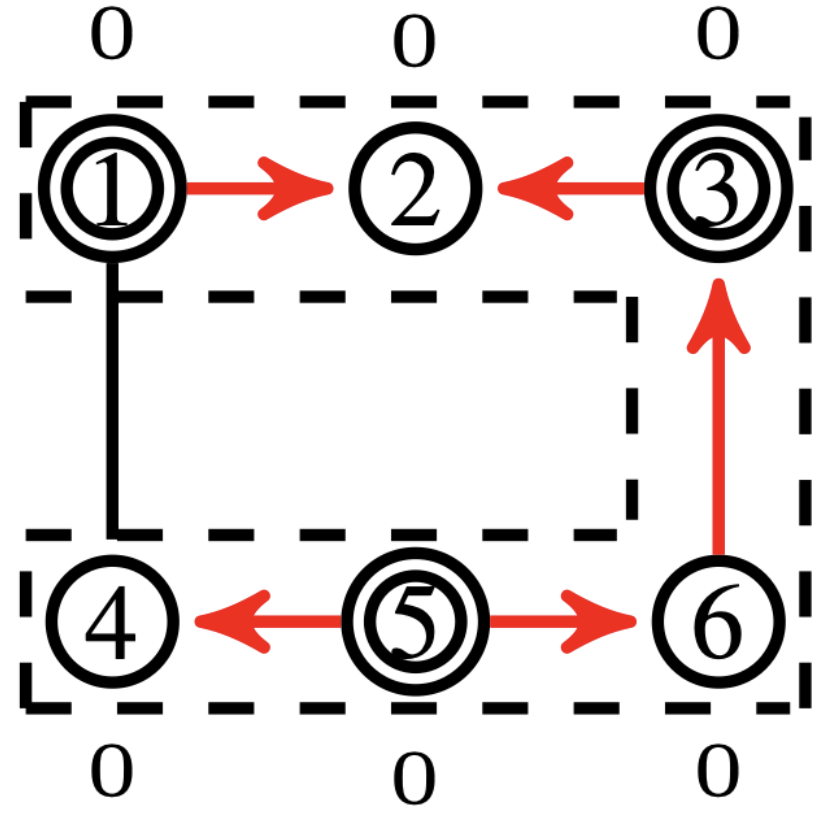} \\
        \small (d) $3^{\textup{rd}}$ step & \small (e) $4^{\textup{th}}$ step & \small (f) $5^{\textup{th}}$ step 
    \end{tabular}
    \medskip
    \caption{{\small Illustration of the sampling procedure. In b) and c), note that priority queue in \textsf{Sampler} avoids the flow blockage.}}
    \label{fig::feasibility}
\end{figure}

\begin{lem}
    \textsf{FORWARD} radial configuration will contain all nodes from a sub-graph $\mathcal{G}^\ell(\mathcal{V}^\ell,\mathcal{E}^\ell)$, i.e., $\mathcal{V}(\mathcal{S}^\ell)=\mathcal{V}^\ell$. 
    \label{lem::connectivity}
\end{lem}
\begin{proof}
    Given that the \textit{edge-delete} procedure only removes edges between nodes within the sampled polyforest, nodes that have not been connected to any polytree will retain all of their edges in $\mathcal{E}^\ell$. Consequently, a node can only remain isolated if none of the connected polytrees can provide sufficient input flow to satisfy the output demanded by the node. Assume there exists a node $j \in \bar{\mathcal{V}}^\ell$ (node 2 in Fig.~\ref{fig::feasibility}) whose output cannot be supplied by any of its adjacent connected polytrees in $ \mathcal{N}(j)$ (nodes 1, 3 in Fig.~\ref{fig::feasibility}). Since $ \bar{\mathcal{G}}^\ell $ is bi-connected, each super source node $i \in \mathcal{N}(j)$ is connected to super sink nodes distinct from  $j$ (e.g., 1 to 4 and 3 to 6 in Fig.~\ref{fig::feasibility}). To maintain overall flow balance in $\bar{\mathcal{G}}^\ell$, at least one super source node $u$ in the partition must have sufficient input flow to meet the output of its neighbors (node 5 in Fig.~\ref{fig::feasibility}). Moreover, due to the bi-connectedness of $\bar{\mathcal{G}}^\ell$, there exists a super sink node $v$ such that $v \in \mathcal{N}(i) \cap \mathcal{N}(u)$ (nodes 4, 6 in Fig.~\ref{fig::feasibility}). By definition, $|p_{i}^\ell| \leq |p_{v}^\ell| \leq |p_{u}^\ell|$.

    Since $\bar{\mathcal{G}}^\ell$ is quasi-bipartite, if $i$ is not receiving extra input to meet output in $j$, it is because flow from $u$ cannot pass through $w$. This scenario arises only if the edge $(i \rightarrow v)$ is sampled in $\mathcal{S}^\ell$ before $(u \rightarrow v)$, turning $v$ into a ``pseudo" source. For instance, in Fig.~\ref{fig::feasibility}, if \( 1 \rightarrow 4 \) or \( 3 \rightarrow 6 \) were sampled before \( 5 \rightarrow 4 \) or \( 5 \rightarrow 6 \), this contradicts the priority queue \( q \) in \textsf{Sampler}, which prioritizes polytrees with excess input flow with respect to its neighbours (node 5 in Fig.~\ref{fig::feasibility}). Thus, by contradiction, this situation cannot occur, ensuring that \textsf{FORWARD} successfully connects all nodes in \( \bar{\mathcal{V}}^\ell \).
\end{proof}

In light of the aforementioned, we can conclude the demonstration by noting that the solution provided by \textsf{FORWARD} will be a feasible radial configuration for flow distribution.

\section{Case study}

We demonstrated the efficacy of the proposed \textsf{FORWARD} algorithm in solving an optimal power distribution problem and discuss how edge weights (used in line 4 of \textsf{Sampler}, Algorithm~\ref{alg::sampler}) can be selected for this problem considering the physical nature of the flow in power networks. It is shown in the literature,  single-phase direct current (DC) simplifications can be used to approximate the real three-phase alternating current (AC) flow problem, casting the optimal power flow problem as \eqref{eqn::problem1}. We refer the readers to~\cite{russell} for specific details of these simplifications.

In a power network, the demand at the sinks (loads) and the input in the sources (generators) are power. The flow in edges is characterized by the demand of the nodes in each distribution line following Ohm's law. In power network systems, $C_{i,j}$ in the cost function of~\eqref{eqn::problem1} is the resistance $R_{i,j}$ along the edge $(i,j)$.

\emph{Weight design}: Given a radial configuration, the flow $x_{i,j}$ at edge $(i\to j)$ is defined as a function of the power demand $d_{i\to j}$ of all the loads in power line extending from the source and containing edge $(i\to j)$ and the voltage in node $j$. Therefore, the cost due to flow through $(i\to j)$ is
$$f_{(i\to j)} = R_{i,j}\cdot(\frac{d_{i\to j}}{v_j})^2.$$

\textsf{FORWARD} constructs the radial configuration by iteratively adding the edges. Therefore, the power line connecting the nodes is not known in advance. Consequently, to have a cost-aware edge choice, the greedy selection process must estimate $d_{i\to j}$. For power networks we propose to approximate the energy lost due to flow across an edge $(i\to j)$, when we want to sample this edge and add it to an existing radial configuration (power line) $\mathcal{T}^\ell_k$, to use 
\begin{equation}
    \label{eqn::loss_edges}
    \begin{split}
        \hat{f}(i\to j) &= R_{i,j}\cdot p_j^2 + \widetilde{h}(\mathcal{T}^\ell_k),
    \end{split}
\end{equation}
where $p_j$ is the power demand at node $j$ and $\widetilde{h}(i)$ is the accumulated cost of power line $\mathcal{T}^\ell_k$.
Thus, a relevant edge weight (probability) for the random walk could be computed as 
\begin{equation}
    \label{eqn::distribution_prob2}
    w_{i,j} = \frac{p_i}{R_{i,j}\cdot d_j^2 + \widetilde{h}(\mathcal{T}^\ell_k)}
\end{equation}
where $p_i$ is the input vector characterized as the power remaining in node $i$ which decreases in each step by subtracting the power demand on $j$. Therefore, the weights on each edge are dynamically changing in function of the steps taken. We point out here that of use of transition probability for \eqref{eqn::distribution_prob2} is with a degree of abuse of notation as in function of the parameters $w_{i,j}$ can be greater than $1$. Indeed, $w_{i,j}$ is normalized at each iteration considering all the edge candidates to be sampled at time $t$.


\emph{Numerical demonstration}: In order to show computational benefit  of using \textsf{FORWARD} in producing a feasible radial configurations for power network  problems and the optimality gap of the algorithm, we consider $6$ different power distribution networks.  Specifically, three IEEE graphs have been used: IEEE $13$ with $2$ sources, IEEE $18$ with $2$ sources and IEEE $33$ with $3$ sources. Note that, as mentioned before, these IEEE graphs have been pre-processed to a single-phase DC network structure. Besides, to test scalability, three random distribution systems have been generated following Watts–Strogatz mechanism \cite{Watts1998} in order to create small-world graph structures. These graphs are named: WS $120$, WS $240$ and WS $400$, where their number indicates the number of nodes in each network. Networks WS $120$ and WS $240$ where designed with $10$ sources and WS $400$ has $20$ source nodes. We have modeled our numerical problem in PowerDistributionModel (PMD) framework~\cite{PMD} from Los Alamos National Laboratory, which is coded in Julia. This framework uses Knitro, from Artelys, to solve the MINLP. This numerical study was conducted on a MacBook Air with an M3 chip and $24$ GB of RAM. The simulation results are shown in Table~\ref{tab:sample_table}. In this table, power loss is in giga-watts (GW). In time column, $*$ stands for processes which has been manually stopped after certain CPU Time, and $-$ stands for situations where a solution could not be found after $3$h of computation.

According to Table~\ref{tab:sample_table}, Knitro was only able to return a solution for small-sized networks. For these cases, the CPU time of Knitro is significantly more than the CPU time used to implement \textsf{FORWARD}. Knitro (like other commercial solvers) uses heuristics to warm up the process within the feasible domain. Notably, as the network size increases, these heuristics tend to struggle to find a proper initialization within polynomial time due to the large dimension of the combinatorial space. This often results in prolonged CPU times or even the inability to find an initial point. In contrast, \textsf{FORWARD} was able to construct a feasible radial configuration for all cases in noticeably low time, even for a network of 400 nodes.

The results shown in Table~\ref{tab:sample_table} for the cost indicate that \textsf{FORWARD} attains a promising optimality gap. This could be attributed to the physically flow-aware nature of designing the weights and the sampler mechanism of \textsf{FORWARD}. In our future work, we plan to formally characterize the optimality gap of \textsf{FORWARD}.



\label{sec:numeric}
\begin{table}[t]
\centering
\caption{Comparison between Knitro and \textsf{FOWARD}. }
\label{tab:sample_table}
\begin{tabular}{|c||c|c||c|c|}
\hline
~&\multicolumn{2}{c||}{Knitro} &\multicolumn{2}{c|}{\textsf{FORWARD}} \\\hline
Graph & cost  & CPU time & cost & CPU time \\ 
\hline
IEEE 11 & 1.19e-7 & 118.21 & 1.19e-7 & 1.81 \\ \hline
IEEE 18 & 1.79e-7 & 141.52 & 1.79e-7 & 1.53 \\ \hline
IEEE 33 & 3.18e-09  & 500* & 9.19e-09  & 5.2 \\ \hline
WS 120 & - & - & 1.42e-11 & 26.91 \\ \hline
WS 240 & - & - & 4.39e-9 & 130.63 \\ \hline
WS 400 & - & - & 2.83e-9 & 217.72 \\ \hline
\end{tabular}
\end{table}

\section{Conclusions}
\label{sec:conclusions}
We presented an algorithm for finding a feasible radial reconfiguration in distribution networks. Promising results from numerical experimentation allows us to believe the proposed method can serve as an effective warm-up strategy for iterative solvers refining solutions toward optimality. Also, it should be emphasized that despite the algorithm is designed for a simplified transportation problem, it can be seen it handles extended problems with complex physics constraints as they are power systems reconfiguration. For future work, it will be shown how, indeed, proposed problem works as an abstraction over other problems with non-linear constraints. 


\bibliographystyle{ieeetr}

\end{document}